\documentclass{ws-mplb}

\begin{document}
\title{POINT-CONTACT SPECTROSCOPY OF MAGNESIUM DIBORIDE WITH DIFFERENT COUNTER-ELECTRODES}

\author{A. I. D'YACHENKO$^\dag$, V. YU. TARENKOV$^\dag$, M. A. BELOGOLOVSKII$^\dag$, V.~N.~VARYUKHIN$^\dag$,
A. V. ABAL'OSHEV$^\ddag$\footnote{Corresponding author.} \ AND S. J. LEWANDOWSKI$^\ddag$}

\address{$^\dag$Donetsk Physical and Technical Institute, National Academy of Sciences of
Ukraine, R. Luxemburg 72, 83114 Donetsk, UKRAINE}
\address{$^\ddag$Instytut Fizyki Polskiej Akademii Nauk, Al. Lotnik\'{o}w 32, 02-668 Warszawa,
POLAND\\
abala@ifpan.edu.pl}

\maketitle

\begin{history}
\received{\today}
\end{history}

\begin{abstract}
We report on tunneling and Andreev-reflection conductance spectra of $39{\:}$K
superconducting magnesium diboride, obtained with Pb and Au counter-electrodes. Two
distinct steps at close to $2.7$ and $7.1{\:}$meV appear in a low-resistance metallic-type
Au-MgB$_2$ junction characteristic, whereas a tunneling-like spectrum measured for the
same junction, annealed by the application of dc current, exhibits only a rounded
contribution of the larger gap. Junctions with a superconducting lead counter-electrode
pressed into a bulk MgB$_2$ sample reveal two conductance peaks that are interpreted as
the result of the formation of highly-transmitting break junctions inside the magnesium
diboride ceramic. Our results strongly support the two-band model with two different gap
values on quasi-two-dimensional $\sigma $ ($7.1{\:}$meV) and three-dimensional $\pi $
($2.7{\:}$meV) Fermi surface sheets of MgB$_2$.
\\ \\
PACS Number(s): 74.25.Kc, 74.45.+c, 74.50.+r, 74.70.Ad
\end{abstract}

\section{Introduction}
The discovery of superconductivity in the simple binary compound, MgB$_2$, generated
an intense research activity  -- mainly, because MgB$_2$ exhibits a transition
temperature $T_c$ of $39{\:}$K, which is almost two times higher than that of any
other intermetallic superconductor known earlier~\cite{Buzea}, but also because of
good perspectives for practical applications. During the two years, which have
passed from this discovery, our understanding of physical properties of the novel
superconductor has made rapid progress.

After confirmation~\cite{Kotegawa} of the spin-singlet type of the Cooper pairs
involved, the main discussions relating to fundamental physics concentrated on two
issues: the mechanism of  the Cooper pair formation and the symmetry of the order
parameter. The boron isotope effect experiments~\cite{Bud'ko} were the first
indication of phonon-mediated superconductivity. Direct experimental probes of
phonon-induced structures in point-contact~\cite{Bobrov,Naidyuk,Yanson,Samuely} and
tunneling~\cite{Dyachenko} superconducting junctions have provided unequivocal
support to the theories~\cite{An,Kong,Bohnen,Liu,Choi1,Choi2}, which explain the
relatively high $T_c$ as originating from the strong anharmonic electron-phonon
coupling to Raman-active high-frequency $E_{2g}$ vibration modes, corresponding to
the in-plane distortions of the boron hexagons.

It took much longer to elucidate the properties of the superconducting order parameter,
but now from the results of tunneling and photoemission spectroscopy, specific heat, and
other measurements emerged strong experimental evidence that magnesium diboride belongs to
a rare class of materials with two gaps of different widths, which arise from a huge
anisotropy of electronic properties ({\it cf} the review of tunneling data for magnesium
diboride~\cite{Schmidt}). The Fermi surface of MgB$_2$ is very anisotropic and consists of
four bands: two $\sigma$-type quasi-two-dimensional cylindrical hole sheets and two
$\pi$-type three-dimensional tubular networks, exhibiting very different electron-phonon
coupling strengths~\cite{Kortus}. As a result, the superconducting gap varies considerably
on the Fermi surface and, as follows from first-principles' calculations~\cite{Choi2},
clusters into two groups. The $\sigma$ cylindrical sheets relate to a larger gap of an
average value of $7{\:}$meV with small variations, while the smaller gap ranges from $1.2$
to $3.7{\:}$meV on the three-dimensional $\pi $ sheets. These calculations support a
two-gap scenario for magnesium diboride, originally proposed by Shulga {\it et
al.}~\cite{Shulga} in order to explain the behavior of the upper critical magnetic field.
It should be noted that, in contrast to conventional superconductors, impurity scattering
does not average out strongly different gap values. Averaging becomes ineffective in
MgB$_2$ because the different symmetry of $\sigma$- and $\pi$-type bands makes interband
scattering much weaker than intraband impurity mixing of electron states~\cite{Mazin}.

Based on such ideas, an effective two-band model with $\Delta _{\sigma }(T=0) =
7.1{\:}$meV and $\Delta _{\pi }(T=0) = 2.7{\:}$meV was derived, and consequences for
tunneling into MgB$_2$ were deduced~\cite{Brinkman}. The discussion of our tunneling and
point contact data relating to the energy gap distribution will be mainly based on this
model and will use a minimal number of fitting parameters. Most samples studied previously
were in the intermediate regime between a direct metallic contact and a tunnel junction
with a strong insulating (I) barrier. The main conclusions for these structures have been
drawn by fitting the differential conductance data to the one-dimensional
Blonder-Tinkham-Klapwijk (BTK) model~\cite{BTK}, modified by introducing a damping
parameter $\Gamma$, and by taking into account two possible channels for the charge
transport, so that not only the gap values, but also the ratio of the individual channel
contributions were taken as adjustable parameters. In this paper, we present the data for
low-resistance junctions with a gold counter-electrode, which to our knowledge are the
first reported direct metallic-type contacts with superconducting MgB$_2$, and which can
be described within the two-band model~\cite{Brinkman} without any additional parameters.
Moreover, by applying dc current pulses we have successfully transformed the clean contact
into a high-resistance tunneling junction, and reproduced in this way the two limiting
regimes of the theory~\cite{Brinkman}. The second issue, which is discussed in the paper,
is the assertion by Schmidt {\it et al.}~\cite{Schmidt,Schmidt1} that pressing a
counter-electrode into the MgB$_2$ sample often breaks off an MgB$_2$ crystal fragment and
results (even if we are dealing with a normal (N) injector) in a junction between two
(identical) superconductors, instead of the expected normal metal~-~superconductor
structure. Such internal break-junctions apparently can be formed also when a
superconducting counter-electrode is pressed into MgB$_2$. Our experiments with a lead
injector provide an independent confirmation of the observation made by Schmidt {\it et
al.}, and fit into the double-gap scenario as well.

\section{Results and Discussion}
\subsection{Sample preparation}
We used polycrystalline MgB$_2$ samples in the form of $15\times 1\times 0.1{\:}{\rm
mm^3}$ rectangular bars of high density, prepared by compacting commercial magnesium
diboride powder at $\sim 20-30{\:}$kbar. Magnetization measurements have shown a
superconducting transition with the onset and midpoint at $39.0$ and $36.5{\:}$K,
respectively. All experiments were performed at $4.2{\:}$K. The junctions based on
these samples exhibited mostly no 'native' barrier, and their resistance varied from
several to a few tens of Ohms.  In order to establish a suitable barrier between the
electrodes, we applied short current pulses with an amplitude of several tens of
milliamperes and a duration of a few seconds; as a result we obtained junction
resistances up to a few hundred Ohms. Two different counter-electrodes have been
used: lead and gold. Current $I$ versus voltage $V$ characteristics as well as
differential conductance spectra $G(V)=dI(V)/dV$ were measured by conventional
four-probe lock-in technique~\cite{Wolf}. In the paper we present typical data
obtained in this manner.

The junctions with Pb injector were fabricated by mounting a Pb hemisphere of
$0.5{\:}$mm diameter on an elastic beryllium strip, which was driven into the
MgB$_2$ surface. Point-contact junctions with an Au counter-electrode were prepared
by touching the superconducting surface with a $0.02$~mm gold wire wrapped around
the MgB$_2$ sample; the formation of SNS junctions was avoided in this manner. The
device was stabilized by a coat of varnish. This procedure, which yielded junction
resistance of about $10{\:}\Omega$ and conductance spectra indicating a direct
heterostructure formed by normal and superconducting electrodes, could in principle
produce several point contacts instead of a single one. Such occurrence is not
important in the case of clean junctions, since in such a case the spectral
contribution of the contacts with transparency significantly smaller than unity is
negligible. In our best Au-MgB$_2$ samples, an almost ideal metallic Sharvin contact
between gold and magnesium diboride was formed. This can be seen from the
conductance drop with voltage increase from $V=0$ to $\Delta _\sigma /e$
(figure~\ref{f.1_Au}). The drop is almost by a factor of two, with the small
deviation attributed to damping and thermal smearing effects.

\subsection{Pb-MgB$_2$ junctions}
\begin{figure}[!htbp]
\centerline{\psfig{file=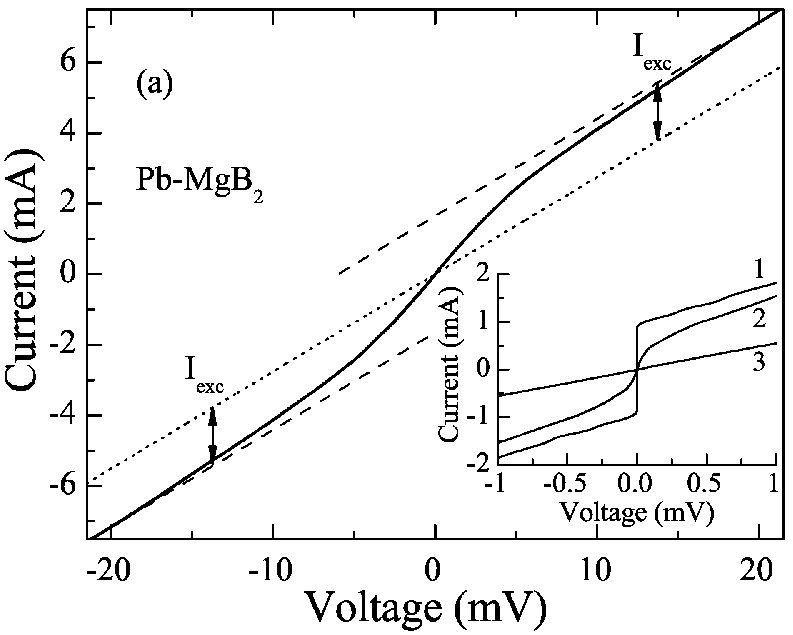}} \vspace{0.5cm} \centerline{\psfig{file=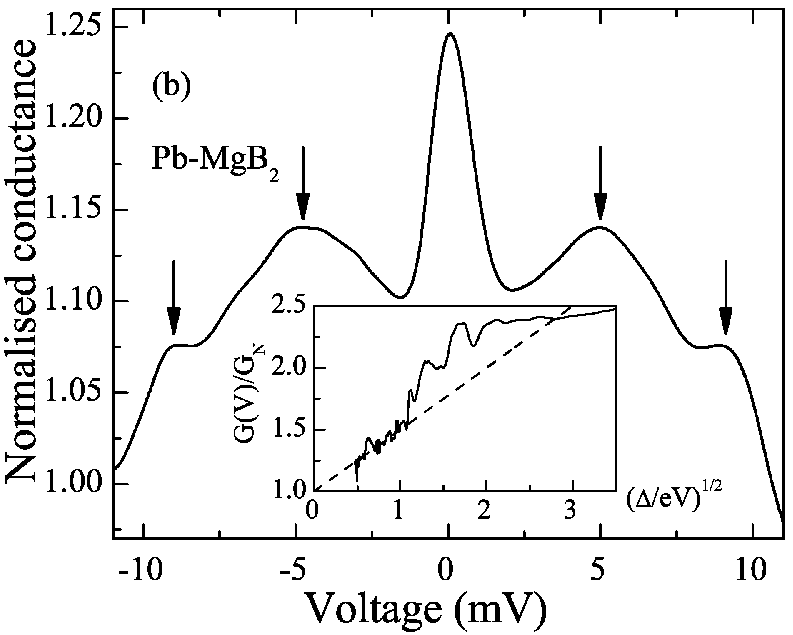}}
\caption{Current and differential conductance in function of voltage for a junction formed
by pressing Pb into MgB$_2$ bulk sample: \ (a) $I-V$ characteristic at $T=4.2\:$K of a
current modified junction (marked $3$ in the inset) exhibiting an excess current $I_{{\rm
exc}}=1.6\:$mA (solid line in the main panel); the dotted line was measured in the normal
state. The inset shows successive modifications of the $I-V$ characteristics (labelled $1$
to $3$) after applying short current pulses. \ (b) differential conductance normalized to
the conductance at $20\:$mV corresponding to characteristic $3$ demonstrates a prominent
zero-bias maximum with two peaks on each side (arrows indicate voltage positions of the
combinations of small and large gap values $2\Delta_\pi/e$ and $\Delta_\sigma/e+\Delta_\pi
/e$). The inset shows $G(V)/G_{{\rm N}}$ corresponding to characteristic $2$ as a function
of $\sqrt{\Delta/{eV}}$ (continuous line) and the plot of the
theoretical~\protect\cite{Bardas} interpolating formula for disordered SNS junctions
$G(V)/G_{{\rm N}}=1+0.5\sqrt{\Delta/eV}$ (dashed line); the best fit was found for
$\Delta=6.8\:$meV.} \label{f.1_Pb}
\end{figure}

The junctions with a Pb injector were prepared in an analogous manner to that used by
Schmidt {\it et al.}~\cite{Schmidt1}, who formed symmetric ${\rm MgB_2-MgB_2}$ structures
by breaking off an MgB$_2$ crystal fragment under the tip pressure. Following the
arguments of the cited work, we also suppose that our Pb-MgB$_2$ samples are symmetric
break junctions inside the magnesium diboride bulk, rather than contacts between two
different superconductors. The nature of the junction barrier is unknown, but its small
resistance compared to that of the SIS structures~\cite{Schmidt1} suggests that we are
dealing rather with SNS contacts (see also the paper of Gonnelli {\it et al.}~\cite{Gon}
and arguments given below). The initial $I(V)$ characteristic exhibited a critical
supercurrent at zero bias $I_{{\rm c}}\sim 0.9{\:}$mA and normal-state resistance $R_{{\rm
N}}=1.8{\:}\Omega$, i.e., $I_{{\rm c}}R_{{\rm N}} \sim 1.6{\:}$mV [plot $1$ in the inset
in figure~\ref{f.1_Pb}(a)]. This value of $I_{{\rm c}}R_{{\rm N}}$ is significantly lower
than that predicted by the model~\cite{Brinkman} for a Josephson current across  a
symmetric ${\rm MgB_2-MgB_2}$ device, but is consistent with $I_{{\rm c}}R_{{\rm N}}$
values found previously by two other groups~\cite{Gon,Schmidt1} for break junctions of
magnesium diboride.

The evolution of the interface properties of the junctions, caused by the application of
the forming current pulses, is clearly visible in the current-voltage characteristics
shown by the inset in figure~\ref{f.1_Pb}(a). After two successive current pulses
(characteristics `2' and `3' in the inset), we obtained a finite resistance $R_{{\rm
N}}=3.6{\:}\Omega$ at zero bias (characteristic `3'). The latter characteristic, shown in
the main panel of figure~\ref{f.1_Pb}(a) in a larger voltage scale, exhibits an excess
current $I_{{\rm exc}}$ known to be the manifestation of Andreev reflections in
high-transparency superconducting heterostructures. Remarkably, only twofold increase of
junction resistance was sufficient to suppress the supercurrent, whereas the Josephson
current in reference~\cite{Schmidt1} was visible up to contact resistances of about
$10{\:}{\rm k}\Omega$. Our result can indicate that we create multiple
incoherent-scattering centers and the Josephson current is cancelled because of the
phase-breaking events. Liao {\it et al.}~\cite{Liao} found that magnesium diboride samples
prepared by solid-state reaction contain large amounts of precipitates composed mainly of
magnesium and oxygen. We believe that their presence is the origin of a strong damping
factor in conductance spectra of MgB$_2$ (even without any current treatment), stressed in
almost all publications on point-contact spectroscopy of this compound. It is highly
probable that by applying current pulses, we create such centers and destroy the phase
coherence during the charge transfer across the interface region between two
superconducting banks.

To further verify the hypothesis of the SNS symmetric break-junctions formed inside
MgB$_2$ bulk, let us analyze the high-voltage asymptote of the current-voltage
characteristic shown in figure~\ref{f.1_Pb}(a) and known to be determined by a set of
interlayer transparencies~\cite{Naveh}. Their distribution has been found in two extreme
cases: for a diffusive normal conductor by Dorokhov~\cite{Dorokhov} and for a disordered
insulating interface by Schep and Bauer~\cite{Schep_Bauer} (see also Naveh {\it et
al.}~\cite{Naveh}). For two identical superconductors  both theories predict $I(V)=
G_{{\rm N}}(V+k\Delta /e)$, where the last term is the excess current $I_{{\rm exc}}$. For
the Dorokhov distribution, the coefficient $k=1.467$, while within the Schep-Bauer
approximation, $k=1.055$ \cite{Naveh}. Applying the arguments of reference \cite{BTK}, we
presume that for dissimilar superconductors with gap values $\Delta $ and
$\Delta^{\prime}$ the excess current in a SNS$^{\prime }$ junction is given by $I_{{\rm
exc}}R_{{\rm N}}\approx k(\Delta +\Delta^{\prime})/(2e)$. We can perform then an analysis
similar to that presented in reference ~\cite{Brinkman} for the dc Josephson current. For
a symmetric ${\rm MgB_2-MgB_2}$ structure we conclude that the $I_{{\rm exc}}R_{{\rm N}}$
product should vary from $\lesssim4{\:}$mV (tunneling along the $c$-axis direction) to
$6{\:}$mV (tunneling within the $a-b$ plane) in the case of a S-diffusive N-S structure,
and between $3{\:}$mV to $5{\:}$mV in the case of a disordered insulating interface. At
the same time, for a contact formed by a conventional superconductor like Pb with a single
gap equal to $1.4{\:}$mV, the $I_{{\rm exc}}R_{{\rm N}}$ value cannot exceed $4{\:}$mV in
any model. From figure~\ref{f.1_Pb}(a) we find $I_{{\rm exc}}R_{{\rm N}}=5.7{\:}$mV, i.e.
a value much greater than the latter limit. This strongly supports the thesis that the
weak link in our structures with lead counter-electrode is not at the interface of the
contacting materials but rather inside the magnesium diboride bulk. Furthermore, the
product $I_{{\rm exc}}R_{{\rm N}}$ exceeds the upper limit for the Schep-Bauer
distribution, providing an argument for the SNS nature of the junction.

To verify further this scenario, let us compare our dc differential conductance with
the predictions of the scattering-matrix theory of multiple Andreev reflections in
short disordered SNS junctions~\cite{Bardas}. Using zero voltage asymptotic behavior
of the dc differential conductance $G(V)=dI(V)/dV$ (equation (14) of reference
\cite{Bardas}), we can check on the background behavior of our conductance spectra.
We apply an interpolating formula $G(V)=G_{{\rm N}}(1+0.5\tanh(\Delta
/2T)\sqrt{\Delta /eV})$, which is valid both at very low bias (the near-zero-bias
$1/\sqrt{V}$ singularity is the hallmark of Andreev-reflection processes occurring
at each junction face), and at very high voltages (in the latter case $G(V)$ is
equal to the normal-state value $G_{{\rm N}}=1/R_{{\rm N}}$). Using $\Delta$ as a
fitting parameter, we found that the best agreement with the experimental data is
achieved for $\Delta =6.8{\:}$meV. The overall behavior of the experimental
normalized conductance $G(V)/G_{{\rm N}}$ corresponding to the experimental data and
plotted in function of $\sqrt{6.8/{eV}}$ is satisfactorily described [see the inset
in figure~\ref{f.1_Pb}(a)] by a simple fitting line $G(V)/G_{{\rm
N}}=1+0.5\sqrt{\Delta /{eV}}$ in the voltage range above $0.5{\:}$mV, where
temperature smearing is not effective. In the small bias region up to $15{\:}$meV,
the conductance exhibits a fine structure, which according to Bardas and
Averin~\cite{Bardas} can be attributed to subharmonic gap singularities, as well as
to the gap features.

Characteristic `3' is more smooth than `2', and the corresponding differential
conductance presented in the main panel of figure~\ref{f.1_Pb}(b) contains a
strongly suppressed zero-bias peak and two maxima. Their voltage positions are very
close to the combinations of small and large gap values $2\Delta _\sigma/e$ and
$\Delta _\sigma/e+\Delta _\pi /e$ expected for the magnesium
diboride~\cite{Brinkman} and shown in figure~\ref{f.1_Pb}(b) by arrows. We should
emphasize that we do not observe any signs of the lead gap. Moreover, the features
discussed above, including the peak at $V=0$, are seen at temperatures exceeding the
$T_{{\rm c}}$ of Pb, and this rules out their possible relation to the lead
counter-electrode.

\subsection{Au-MgB$_2$ point contact junctions}

Let us examine now the conductance spectra of our Au-MgB$_2$ point contacts in the
framework of a double-gap model with fixed gap values and weighing
factors~\cite{Brinkman}. These quantities are found from first-principles' calculations,
therefore the adjustable parameters are reduced to the barrier strength and the damping
factor $\Gamma $. The latter one was introduced into the tunneling density of states by
Dynes {\it et al.}~\cite{Dynes} and is often referred to as the Dynes factor. The barrier
strength is measured by a parameter $Z=\int\limits_0^dH(x)\,{\rm d}x/\hbar v_{{\rm
F}}$~\cite{BTK}, where $H(x)$ is the barrier height, $x$ denotes the coordinate normal to
the interface, $d$ is the barrier width, and $v_{{\rm F}}$ is the $x$-component of the
Fermi velocity. The conductance spectrum $G(V)$ is assumed to be a sum of two single-mode
BTK contributions~\cite{BTK} related to the two $\sigma$- and $\pi$-type bands in
proportions strongly depending on the injection angle. The relevant weighting factors,
which determine the ratio of the two contributions, are proportional to the squared plasma
frequencies and are given in reference~\cite{Brinkman} for the transport within the $a-b$
plane, $G_{a-b}(V)$, and for tunneling in the $c$-axis direction, $G_{c}(V)$. In the
present experiment, the conductance is a mixture of $a-b$ plane and $c$-axis contributions
with unknown coefficients, which can be different for different polycrystalline samples.
Only to get a qualitative estimate of the averaged conductance spectra predicted by the
theory, we compare the measured spectra with the mean characteristic
$G_m(V)=[G_{a-b}(V)+G_{c}(V)]/2$. As it was stated above, we prefer to obtain qualitative
conclusions with a limited number of fitting parameters rather than to introduce weighting
factors with the aim to improve the agreement between theory and experiment.

\begin{figure}[!htbp]
\centerline{\psfig{file=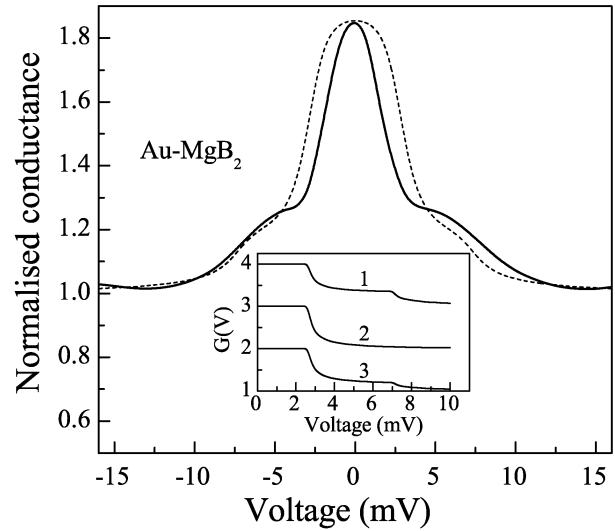}}
\caption{Conductance spectrum at $T = 4.2{\:}$K of a direct Au-MgB$_2$ contact ($R_{{\rm
N}}=20.5{\:}\Omega $), normalized to the conductance at $20{\:}$mV (solid line). The
dashed line represents a mean characteristic $G_m(V)$ calculated from the two-gap
model~\protect\cite{Brinkman} in the limit of a completely transparent interface and $Z=0$, small
smearing parameter $\Gamma = 0.23$ and $T = 4.2{\:}$K. The inset shows zero-temperature
theoretical plots of conductance ($Z = 0$, $\Gamma = 0$): $1$ -- charge transport within
the $ab$-plane, $2$ -- transport along the $c$-axis, $3$ -- mean conductance $G_m(V)$; the
plots are displaced vertically for clarity.} \label{f.1_Au}
\end{figure}

In reference~\cite{Brinkman}, a two-step structure was predicted for a direct
contact between a normal-metal injector and magnesium diboride and for charge
transport in the $a-b$ plane (upper curve in the inset in figure~\ref{f.1_Au}
calculated for $T=0$). For the $c$-axis transport, the main contribution comes from
the $\pi$ band (curve $2$ in the inset). The averaged characteristic (see curve $3$
in the inset and the dashed curve in the main panel), which takes into account a
small Dynes factor $\Gamma$ and temperature smearing effects determined by a
standard procedure involving Fermi distribution functions, qualitatively well
reproduces the two-step feature observed by us experimentally (solid line in
figure~\ref{f.1_Au}). It should be stressed that this good agreement of experiment
and theory was obtained with a single fitting parameter $\Gamma$.

\begin{figure}[!htbp]
\centerline{\psfig{file=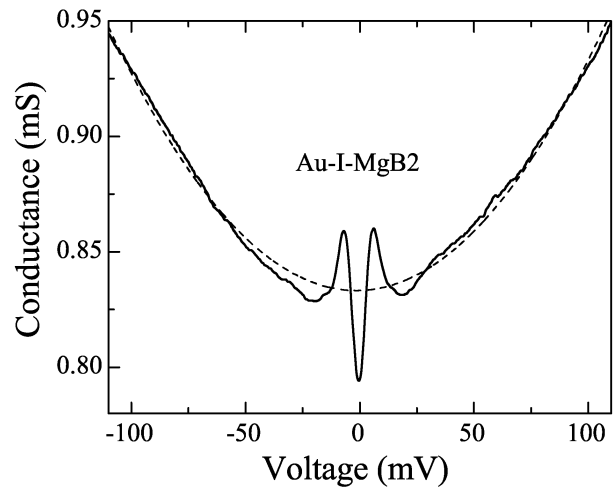}}
\caption{Conductance at $T=4.2{\:}$K of a tunneling Au-I-MgB$_2$ junction with
normal-state resistance $R_{{\rm N}}=1.2{\:}{\rm k}\Omega$ formed by applying several
current pulses (solid line). The dashed line shows a parabolic fit.} \label{f.2}
\end{figure}

We formed next a low-transparency insulating (I) barrier in our Au-I-MgB$_2$ junctions by
applying additional current pulses and measured the resultant $G(V)$ characteristics in a
wide range of bias voltages (up to $100{\:}$meV). The conductance background usually
exhibited a roughly parabolic dependence on voltage, with a very small offset of minimum
from $V= 0$ (see figure~\ref{f.2}), indicating an almost symmetrical barrier~\cite{Wolf}
and proving the tunneling character of the charge transport. A simple
formula~\cite{Simmons} for a rectangular potential barrier may be used to find an average
barrier height $H$ and its thickness $d$. In order to carry out the evaluation, we need
the contact area, which is unknown. Nevertheless, a minimal value for $Z$ parameter can be
estimated from the parabolic shape of the conductance curve. According to
reference~\cite{Simmons}, the coefficient $\kappa $ in the quadratic term in the voltage
dependence of conductance, $G(V)/G(0)=1+({eV}/\kappa )^2$, is given by
$\kappa=H^{1/2}/(0.182d)$, where $H$ is in volts and $d$ is in {\AA}ngstroms. We assume
$d$ to be not less than two-three lattice parameters and take as a minimum $d=10{\:}$\AA .
Substituting $\kappa $ found from the measured conductance background (figure~\ref{f.2}),
we obtain $Z>0.85$. Consequently, we take the first fitting parameter $Z$ equal to $0.9$.

\begin{figure}[!htbp]
\centerline{\psfig{file=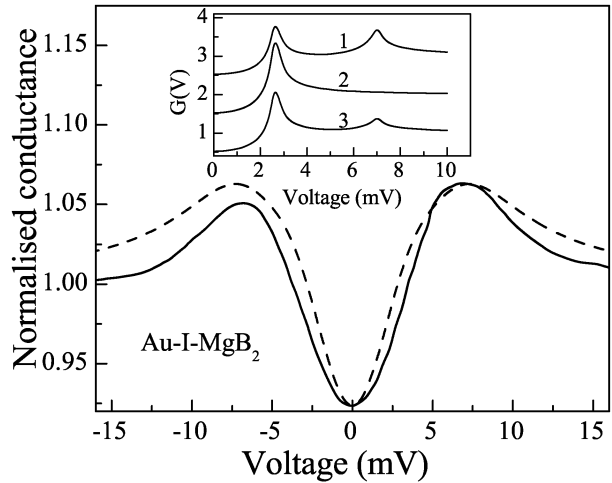}}
\caption{Tunneling-like conductance spectrum of Au-I-MgB$_2$ junction with normal-state
resistance $R_{{\rm N}}=1.2{\:}{\rm k}\Omega$ normalized to the conductance at $20{\:}$mV
(solid line). The dashed line represents the calculated mean characteristic $G_m(V)$
following the two-gap model~\protect\cite{Brinkman} for $Z=0.9$, $\Gamma =2.9$, $T=4.2{\:}$K. The
inset shows zero-temperature theoretical curves for $Z = 0.9$ and vanishing damping
parameter $\Gamma $: $1$ -- charge transport within the $ab$-plane, $2$ -- transport along
the $c$-axis, $3$ -- mean characteristic $G_m(V)$. The plots in the inset are displaced
vertically for clarity.} \label{f.3}
\end{figure}

In figure~\ref{f.3}, the main panel presents the measured conductance spectrum of
the same junction in the near-gap voltage region, while in the inset we show the
conductance calculated for tunneling into the $a-b$ plane and along the $c$-axis
direction for vanishing Dynes factor, $\Gamma =0$. The latter characteristics
clearly indicate that the most pronounced feature in the average conductance
spectrum should be the smaller gap peak, whose signature is missing from the
measured spectrum. The most probable cause is the strong disorder induced by the
forming current pulses and the resulting significant inelastic scattering near the
interface. For large values of $\Gamma$ (the second fitting parameter) the agreement
between the experimental and calculated conductances is very good, as shown in the
main panel in figure~\ref {f.3}.

To conclude the discussion of near-gap features in the junction characteristics, let us
make two remarks concerning the role of disordering effects. The analysis based on a
simple BTK approach~\cite{BTK} for describing the experiment with (probably) numerous
elastic and inelastic scatterers randomly distributed within the barrier appears to be
somewhat naive, nevertheless is justified, since one can always introduce an effective
transparency parameter $Z$ within a single-mode model~\cite{BTK} by integrating over a
certain distribution of transparencies (see the corresponding discussion in
reference~\cite{Bardas}). It should be stressed that the additional averaging already
takes into account the possible appearance of multiple point contacts in our experiments.
The averaging provides some additional smearing, but the most transparent contacts will
dominate in the spectrum.

The second remark concerns the possibility of observing double-gap spectrum in a
disordered superconductor with important charge scattering. Our samples contain
comparatively large grains. Therefore, we expect that the measured characteristics of a
clean point contact reflect the bulk properties of the material. It is known that dense
bulk samples of MgB$_2$ are usually in the clean limit, with the elastic mean free path
large in comparison to the superconducting coherence length. This condition is necessary
to observe two distinct gaps in a sufficiently clean superconductor. At the same time, our
method of fabricating the insulating layer between the electrodes generates, as it was
argued above, a great amount of inelastic-scattering inclusions. At first sight, the
interband impurity scattering should cause the gaps to converge. But this is not the case
for magnesium diboride. According to reference~\cite{Kuz}, optical properties of dense
polycrystalline samples of MgB$_2$, as well as a surprisingly small correlation between
the defect concentration and $T_{{\rm c}}$, can be understood only by assuming negligibly
small inter-band impurity scattering. The possible reason is in the specific electronic
structure of MgB$_2$, namely the the two-dimensional $\sigma $-band states overlap
slightly with the states in the Mg plane, where defects are most likely to
occur~\cite{Mazin}. Thus one can expect that the interband scattering rate between the
$\sigma $- and $\pi $-bands is small, even in low-quality samples.

\section{Conclusions}
In summary, we have realized several different types of heterostructures based on
superconducting magnesium diboride: from a low-resistance (almost ideal) Au-MgB$_2$ point
contact to an SNS junction (with a lead counter-electrode) obtained by applying current
pulses. Moreover, with this technique we have succeeded to transform point-contact samples
with gold injector into tunnel-like devices. In the case of lead electrode, we have
confirmed the observation of Schmidt {\it et al.}~\cite{Schmidt} that pressing the
electrode can result in the formation of symmetric superconductor-superconductor junctions
inside the magnesium diboride bulk. Our findings for Au and Pb counter-electrodes were
interpreted within the two-gap model for the energy spectrum of MgB$_2$ and thus provide
novel arguments in favor of a multi-band scenario for magnesium diboride.

\section*{Acknowledgements}
This work was partially supported by the Polish Committee for Scientific
Research (KBN) under Grant No. PBZ-KBN-013/T08/19 and by INTAS under project No.
2001-0617.

\end{document}